\documentclass[pra,aps,reprint,showpacs]{revtex4-1}
\usepackage{amssymb}
\usepackage{amsfonts}
\usepackage{amsmath}
\usepackage{graphicx}
\usepackage{epstopdf}
\usepackage{color}

\providecommand{\U}[1]{\protect\rule{.1in}{.1in}}

\begin{document}

\title{Interactions of three-dimensional solitons in the cubic-quintic model}
\author{Gennadiy Burlak$^{1}$ and Boris A. Malomed$^{2}$}
\affiliation{$^{1}$Centro de Investigaci\'{o}n en Ingenier\'{\i}a y Ciencias Aplicadas,
Universidad Aut\'{o}noma del Estado de Morelos, Cuernavaca, Mor., M\'{e}xico}
\affiliation{$^{2}$Department of Physical Electronics, School of Electric Engineering,
Faculty of Engineering, and Center for Light-Matter Interaction,
Tel Aviv University, Tel Aviv 69978, Israel}
%\keywords{one two three}

\begin{abstract}
We report results of a systematic numerical analysis of interactions between
three-dimensional (3D) fundamental solitons, performed in the framework of
the nonlinear Schr\"{o}dinger equation (NLSE) with the cubic-quintic (CQ)
nonlinearity, combining the self-focusing and defocusing terms. The 3D NLSE
with the CQ terms may be realized in terms of spatiotemporal propagation of
light in nonlinear optical media, and in Bose-Einstein condensates, provided
that losses may be neglected. The first part of the work addresses
interactions between identical fundamental solitons, with phase shift $%
\varphi $ between them, separated by a finite distance in the free space.
The outcome strongly changes with the variation of $\varphi $: in-phase
solitons with $\varphi =0$, or with sufficiently small $\varphi $, merge
into a single fundamental soliton, with weak residual oscillations in it (in
contrast to the merger into a strongly oscillating breather, which is
exhibited by the 1D version of the same setting), while the choice of $%
\varphi =\pi $ leads to fast separation between mutually repelling solitons.
At intermediate values of $\varphi $, such as $\varphi =\pi /2$, the
interaction is repulsive too, breaking the symmetry between the initially
identical fundamental solitons, there appearing two solitons with different
total energies (norms). The symmetry-breaking effect is qualitatively
explained, similar to how it was done previously for 1D solitons. In the
second part of the work, a pair of fundamental solitons trapped in a 2D
potential is considered. It is demonstrated that they may form a slowly
rotating robust \textquotedblleft molecule", if aninitial kicks are applied
to them in opposite directions, perpendicular to the line connecting their
centers.
\end{abstract}

\maketitle

\textbf{The self-focusing cubic nonlinearity is ubiquitous in a large
variety of physical media (optics, plasmas, Bose-Einstein condensates,
etc.), helping to build solitons in these media. However, while the
one-dimensional (1D) solitons are completely stable, they are subject to
destructive instabilities in 2D and 3D, induced by the critical and
supercritical collapse, which is driven by the same cubic nonlinearity in
the multidimensional settings. A well-known solution of the stabilization
problem for 2D and 3D solitons is the addition of a self-defocusing quintic
term, which arrests the collapse. This term naturally appears in models of
optical media (in particular, colloidal suspensions of metallic
nanoparticles). Once the stabilization of the solitons is secured, the next
natural step is to consider interactions between them, which is the subject
of the present work, for the most interesting case of 3D solitons. We
perform the analysis by means of systematic numerical simulations and some
analytical approximations. First, we consider interactions between identical
solitons, with phase shift }$\varphi $\textbf{\ between their complex wave
functions, assuming that the solitons are separated by a relatively small
distance. The outcome of the interaction strongly depends on }$\varphi $%
\textbf{. First, attraction between in-phase solitons, with }$\varphi =0$%
\textbf{, leads to their quick merger into a single soliton, in an almost
fundamental form (with weak residual intrinsic oscillations in it, which is
different from the outcome of the interaction in 1D, where in-phase solitons
merge into a strongly excited \textit{breather}). On the other hand, the
solitons with }$\varphi =\pi $\textbf{\ interact repulsively, separating
from each other. A noteworthy outcome is produced by the interaction in the
intermediate case, with }$\varphi =\pi /2$\textbf{: the pair of initially
identical solitons features symmetry breaking between them, producing two
solitons with \emph{different} integral norms. This nontrivial effect is
explained by considering the amplitude and phase structure of the
two-soliton pair (the symmetry breaking is a consequence of a mismatch
between the pair's \textquotedblleft amplitude center" and \textquotedblleft
phase center"). A challenging situation is addressed in the second part of
the work, which deals with a pair of 3D solitons trapped in a common 2D
potential (the third spatial direction remains unconfined). It is found that
the pair may form a sufficiently robust slowly rotating \textquotedblleft
molecule", if appropriate initial conditions are applied. The theoretical
results reported in the paper suggest new possibilities for experiments with
multidimensional solitons.}

\section{Introduction}

Cubic nonlinearity, which represents the Kerr effects, is a ubiquitous
feature of a great variety of optical media. It is commonly known that the
interplay of the cubic self-focusing with the group-velocity dispersion or
spatial diffraction gives rise to effectively one-dimensional (1D) solitons
in the temporal and spatial domains, respectively \cite{KA}. The same type
of the nonlinearity, which represents attractive interaction between atoms
in Bose-Einstein condensates (BECs), helps to create stable effectively 1D
matter-wave solitons \cite{RGH}. A natural extension of these well-known
results is building multidimensional solitons \cite{special-topics} (in
particular, spatiotemporal solitons in optics \cite{review-Wise,update},
alias ``light bullets" \cite{Silberberg}). However, a well-known problem is
that, while formal soliton solutions to two- and three-dimensional (2D and
3D) nonlinear Schr\"{o}dinger equations (NLSEs) can be easily found in the
numerical form, they are completely unstable due to the presence of the
critical (2D) and supercritical (3D) collapse in the same equation.

The simplest possibility to arrest the collapse and stabilize the
multidimensional solitons is using a medium with nonlinear response of the
cubic-quintic (CQ)\ type, which includes a self-defocusing fifth-order term.
The CQ nonlinearity occurs in various optical media \cite{general},
including chalcogenide glasses \cite{chalco}-\cite{chalco3} and organic
materials \cite{organic}. Recently, it was demonstrated that required values
of the cubic and quintic coefficients can be accurately engineered in
colloidal suspensions of metallic nanoparticles \cite{Cid1,Cid2}. As a
result, 2D spatial fundamental solitons have been created experimentally in
a liquid bulk waveguide \cite{Cid3}. In principle, the CQ nonlinearity may
be realized in BEC too, with the quintic term provided by repulsive
three-body collisions in a relatively dense condensate \cite%
{Fatkh,Cipolatti,SKA}, although this interpretation is hampered by the fact
that the three-body interactions contribute to losses in the condensate \cite%
{3-body-loss,Bloch}.

As concerns the theoretical analysis, the stability of isotropic fundamental
solitons in 2D and 3D media with the CQ nonlinearity is obvious \cite{2D}-%
\cite{SKA}. A challenging issue for this model is a possibility of the
existence of stable 2D and 3D solitons with \textit{embedded vorticity}, as
the saturation of the nonlinearity (in particular, provided by the quintic
self-defocusing) does not, generally, secure the stability of vortex
(ring-shaped) solitons against ring-splitting perturbations \cite{Dima}. The
partial stability of 2D solitary vortices with topological charges $S=1,2,3$
was discovered in direct simulations and rigorously investigated in Refs.
\cite{Manolo} and \cite{Bob}, respectively (stability regions for $S>3$
exist too, but they are extremely narrow); see also Refs. \cite{Humberto}
and \cite{Yakimenko}. In 3D, stability regions for toroidal solitons with $%
S=1$ in single- and two-component NLSE systems with the CQ nonlinearity were
found, respectively, in Refs. \cite{9-authors} and \cite{bimodal} (the
stability of vortex solitons in the two-component 2D model was considered in
Ref. \cite{bimodal2D}). Related results were produced by simulations of the
evolution of multi-soliton 2D \cite{cluster2D} and 3D \cite{cluster}
clusters carrying an overall angular momentum, which demonstrate slow merger
or expansion of the cluster \cite{cluster}.

The objective of the present work is to consider interactions of 3D
fundamental solitons in the framework of the NLSE with the CQ nonlinearity,
which is a relevant problem for the above-mentioned physical realizations of
the model. Thus far, soliton-soliton collisions in the CQ model were chiefly
studied in the 1D setting \cite{Lev} (some results were also reported for
interactions of 2D solitons \cite{interaction-2D}). By means of systematic
simulations of the 3D equation, we demonstrate that soliton-soliton
interactions are essentially inelastic: depending on the initial phase
difference, $\varphi $, two identical solitons merge into a single one at $%
\varphi =0$ (on the contrary to the previously known results for the
inelastic interaction of 1D solitons, the merger creates a nearly stationary
fundamental soliton, rather than a strongly excited breather); the solitons
bounce back at $\varphi =\pi $; and \textit{symmetry breaking} takes place
at intermediate values of $\varphi $, producing a pair of separating
fundamental solitons with \emph{unequal energies }(norms). In addition to
that, we also construct a slowly rotating \textquotedblleft two-soliton
molecule" (this concept is known in 1D models \cite{molecule,molecule2}),
placing a 3D soliton pair, with angular momentum imparted to it, in a 2D
trapping potential. The latter result suggests a possibility of the creation
of a such a specific rotating bound state in the experiment.

The rest of the paper is organized as follows. The model is formulated in
Section II. Numerical results for the soliton-soliton interactions in free
space are summarized in Section III, the formation of the rotating ``soliton
molecule" is reported in Section IV, and the paper is concluded by Section V.

\section{The model}

The NLSE for local amplitude $\Psi $ of the electromagnetic wave in an
optical waveguide, or the mean-field wave function in BEC, with the CQ
nonlinearity in the 3D space $(x,y,z)$, and 2D trapping potential $V(x,y)$,
if any, is written, in the scaled form, as
\begin{gather}
i\frac{\partial \Psi }{\partial t}+\left( \frac{\partial ^{2}}{\partial x^{2}%
}+\frac{\partial ^{2}}{\partial y^{2}}+\frac{\partial ^{2}}{\partial z^{2}}%
\right) \Psi  \notag \\
+[(\left\vert \Psi \right\vert ^{2}-|\Psi |^{4})+V(x,y)]\Psi =0.
\label{GPE-U}
\end{gather}%
In terms of the spatiotemporal propagation in optics, evolution variable $t$
is actually the propagation distance, $x$ and $y$ are transverse
coordinates, and $z$ the temporal coordinate (reduced time) \cite{KA}. In
addition to the Hamiltonian, Eq. (\ref{GPE-U}) conserves the integral norm,
which is scaled energy ($E$) in the optical model, or scaled number of atoms
in BEC:

\begin{equation}
E=\int \int \int \left\vert \Psi \left( x,y,z\right) \right\vert ^{2}dxdydz.
\label{energy}
\end{equation}

First we address the free-space case, with $V(x,y)=0$, while the dynamics of
a rotating bi-soliton \textquotedblleft molecule", trapped in the 2D
harmonic-oscillator potential,%
\begin{equation}
V(x.y)=V_{0}\left[ \left( x-x_{c}\right) ^{2}+\left( y-y_{c}\right) ^{2}%
\right] ,  \label{V}
\end{equation}%
where $\left( x_{c},y_{c}\right) $ are coordinates of the center of the
integration domain (see, e.g., Fig. \ref{fig2} below), is considered
separately in Section IV. In the absence of any potential (free space), Eq. (%
\ref{GPE-U}) conserved the linear and angular momenta, together with the
Hamiltonian and $E$. In the presence of the isotropic trapping potential (%
\ref{V}), Eq. (\ref{GPE-U}) still conserved the $z$-component of the angular
momentum.

In terms of optics, stationary solutions for isotropic fundamental solitons
in the free space, with propagation constant $K>0$ (in terms of BEC, $-K$ is
the chemical potential), are looked for in the usual form,%
\begin{equation}
\Psi \left( x,y,z,t\right) =e^{iKt}U\left( r\equiv \sqrt{\left(
x-x_{c}\right) ^{2}+\left( y-y_{c}\right) ^{2}+z^{2}}\right) ,  \label{U}
\end{equation}%
with real radial amplitude function, $U(r)$, satisfying equation
\begin{equation}
KU=\frac{d^{2}U}{dr^{2}}+\frac{2}{r}\frac{dU}{dr}+U^{3}-U^{5},  \label{D}
\end{equation}%
subject to boundary conditions $dU/dr(r=0)=0$, $U(r)\sim \exp \left( -\sqrt{K%
}r\right) $ at $r\rightarrow \infty $.

\begin{figure}[tbp]
\centering\includegraphics[width=0.99\columnwidth]{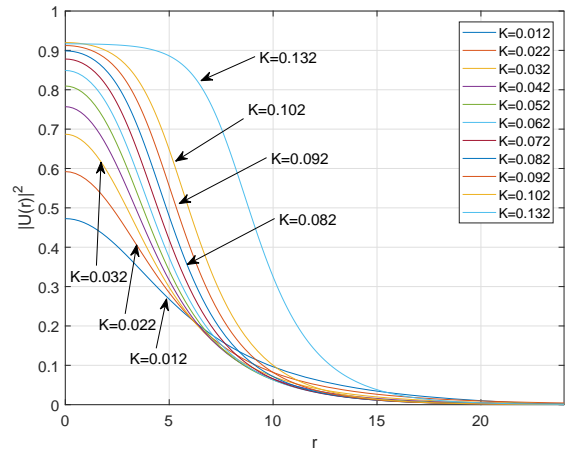}
\caption{(Color online) A set of radial profiles of the 3D isotropic
solitons, found in the numerical form at values of the propagation constant,
$K$, indicated in the figure.}
\label{fig1}
\end{figure}

A set of numerically generated profiles $U(r)$ (actually, they were
obtained, although not displayed, in Ref. \cite{9-authors}) is presented in
Fig. \ref{fig1}. They were produced by a numerical method based on
Newton-Raphson iterations \cite{CppRecipes2002}, implemented in the
Cartesian coordinates in the 3D domain of size $L\times L\times L$, where $%
L=110$ was sufficient to accommodate all the solitons and two-soliton
complexes considered in this work. The stationary solutions were obtained
with relative accuracy $10^{-8}$. Simulations of the evolution of 3D states
were then performed by means of the well-known split-step algorithm \cite%
{Yang:2010a}.

Equation (\ref{D}) generated 3D fundamental solitons at $K<K_{\max }=3/16$,
with $K_{\max }$ determined by the exact soliton solution for the 1D NLSE
with the CQ nonlinear terms \cite{Bulgaria,soliton2}; this happens because
very broad 3D solitons become quasi-one-dimensional in the limit of $%
K\rightarrow 3/16$. In this limit, the solitons develop a flat-top shape
with a diverging radial size, $R\simeq \left( 1/\sqrt{3}\right) \ln \left(
\left( 3-16K\right) ^{-1}\right) $, and the asymptotic value of the
amplitude in the flat-top area, $U\left( K=3/16\right) =\sqrt{3}/2$. As
shown in Ref. \cite{9-authors}, the fundamental 3D solitons are stable at $%
K>K_{\min }\approx 0.02$, hence all the solitons whose profiles are
displayed in Fig. \ref{fig1} are stable, except for the one corresponding to
$K=0.012$.

Lastly, it is relevant to mention that the shape of the fundamental solitons
which do not yet feature an extended flat-top profile, may be accurately
predicted by the variational approximation based on the Gaussian ansatz, $%
U(r)=A\exp \left( -r^{2}/W^{2}\right) $, with constants $A$ and $W$
representing the soliton's amplitude and width. These results are not
displayed here in detail, as the approximation is similar to that developed
previously for the NLSE with the CQ nonlinearity in 1D \cite{De
Angelis,Cipolatti}, 2D \cite{2D}, and 3D \cite{Jovanoski} settings.

\section{Interactions of three-dimensional solitons}

The interaction between two solitons in the framework of Eq. (\ref{GPE-U})
is initiated by input

\begin{gather}
\Psi _{0}\left( x,y,z\right) =\exp (i\varphi /2)U\left( x-x_{c}-d/2\sqrt{2}%
,y-y_{c}-d/2\sqrt{2},z\right)  \notag \\
+\exp (-i\varphi /2)U\left( x-c_{c}+d/2\sqrt{2},y-y_{c}+d/2\sqrt{2},z\right)
,  \label{input}
\end{gather}%
where $U\left( x-x_{c}\mp d/2\sqrt{2},y-y_{c}\mp d/2\sqrt{2},z\right) $ are
the stationary solitons [see Eq. (\ref{U})], whose radial profiles are shown
in Fig. \ref{fig1}, with separation $d$ between their centers [directed
along the diagonal in the $\left( x,y\right) $ plane], and phase shift $%
\varphi $ between them. Generic outcomes of the interaction may be
adequately displayed for soliton pairs with $K=0.062$, separated by distance
$d=10$, by varying phase $\varphi $.

\subsection{The merger of in-phase solitons ($\protect\varphi =0$) and
rebound of out-of-phase ones ($\protect\varphi =\protect\pi $)}

\begin{figure}[tbp]
\centering%{Sol3D_2x2L90t150Phase0.eps}%
\includegraphics[width=0.99\columnwidth]{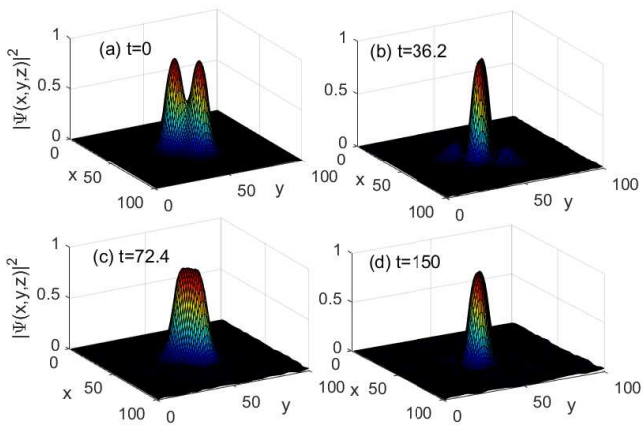}
\caption{(Color online) A typical example of the merger of two in-phase
solitons (with phase shift $\protect\varphi =0$), initially separated by
distance $d=10$ in the diagonal direction, as per Eq. (\protect\ref{input}),
the propagation constant of each soliton being $K=0.062$. In this figure,
and in Figs. \protect\ref{fig5}, \protect\ref{fig6}, and \protect\ref{fig9},
\protect\ref{fig10} below, the density profile is displayed in the midplane,
$z=0$.}
\label{fig2}
\end{figure}

\begin{figure}[tbp]
\centering%{Sol3D_isoL90t150Phase0.eps}%
\includegraphics[width=0.99\columnwidth]{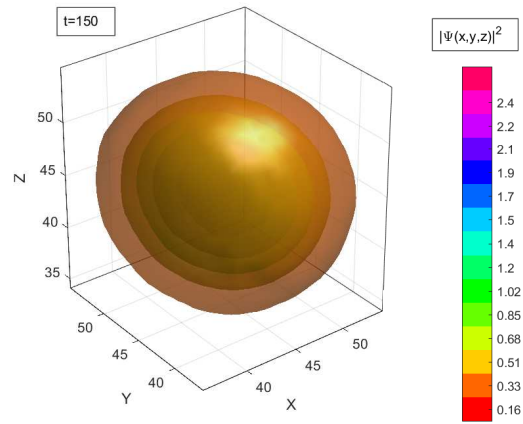}
\caption{(Color online) The shape of the soliton, produced by the merger of
the in-phase soliton pair displayed in Fig. \protect\ref{fig2}, is displayed
by means of isosurfaces of $\left\vert \Psi \left( x,y,z\right) \right\vert
^{2}$ at $t=150$.}
\label{fig3}
\end{figure}

Figure \ref{fig2} shows a typical example of the interaction of two
identical in-phase solitons, with zero phase difference. It is known that
the sign of the interaction between fundamental 3D solitons is attractive
for $\varphi =0$ in input (\ref{input}) \cite{potential}. In accordance with
this, the in-phase solitons attract each other, quickly merging into a
single one. Figure \ref{fig3} demonstrates that the soliton produced by
merger is a nearly isotropic state (although the input is obviously
anisotropic), whose shape is very close to that of the stationary
fundamental soliton with the same value of total energy (\ref{energy}). The
simulations demonstrate residual oscillations of the local power at the
center of the emerging single fundamental soliton, $\left\vert \Psi \left(
x=x_{c},y=y_{c},z=0\right) \right\vert ^{2}$, with a relative amplitude $%
\simeq 10\%$, defined in terms of the density values. The merger occurs as
well at sufficiently small nonzero values of $\varphi $ (in particular, at $%
\varphi =\pi /4$, see Fig. \ref{fig8} below). The intrinsic oscillations may
be subject to gradual damping due to radiation losses, but verification of
this assumption requires running the 3D simulations on a extremely long time
scale, which is a technically challenging objective, and it may correspond
to unrealistically large values of the propagation distance or evolution
time, in terms of the experimental realization in optics or BEC,
respectively. In fact, emission of radiation is observed quite clearly in
oscillations of the breather formed as a result of the merger of 1D solitons,
see Fig. \ref{fig4} below.

In terms of the experimental realization in optics, for 3D solitons
(\textquotedblleft light bullets") with transverse radius $\sim 20$ $\mathrm{%
\mu }$m and carrier wavelength $\sim 1~\mathrm{\mu }$m, which are relevant
values, $t=150$ (the value of $t$ at which the merger has been completed) in
scaled units corresponds, in physical units, to propagation distance $\sim 1$
cm, which is realistic for the experimental implementation.

In-phase solitons governed by the one-dimensional NLSE with the CQ
nonlinearity [Eq. (\ref{GPE-U}) in which terms $\left( \partial
^{2}/\partial y^{2}+\partial ^{2}/\partial z^{2}\right) \Psi $ are dropped]
also interact attractively, which leads to their merger (which is possible
in the framework of the nonintegrable equation), but an essential difference
from the 3D model is that, as shown in Fig. \ref{fig4}, the merger of 1D
solitons produces a nonstationary \textit{breather}, rather than a
nearly-fundamental soliton. The breather is a robust state which persists in
the course of indefinitely long simulations. An explanation to this drastic
difference is that the 1D equation is relatively close to the integrable
cubic NLSE, which has well-known exact solutions in the form of breathers
(higher-order solitons) \cite{Sats-Yaj}. On the other hand, the 3D NLSE with
the CQ nonlinearity is very far from any integrable limit, hence the 3D
equation gives rise to the quick fusion of the colliding fundamental
solitons into a new one, rather than forming a complex mode with strong
inner vibrations.
\begin{figure}[tbp]
\centering
\par
%{Sol1DPhi0Sft3.eps}%
\includegraphics[width=0.99\columnwidth]{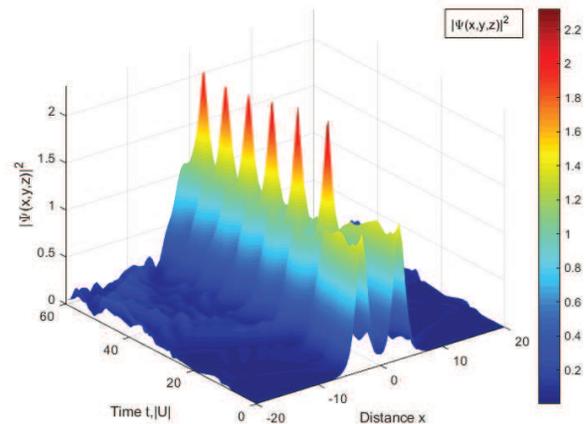}
\caption{A typical example of the merger of two in-pase solitons into a
robust breather with strong intrinsic oscillations, as observed in the
one-dimensional NLSE with the CQ nonlinearity. The initial distance between
the solitons is $d=6$, each one corresponding to the propagation constant $%
K_{\mathrm{1D}}=0.11$. }
\label{fig4}
\end{figure}
{\ }

\begin{figure}[tbp]
\centering%{Sol3D_2x2L90T150LstPhasePiSft5Symm.eps}%
\includegraphics[width=0.99\columnwidth]{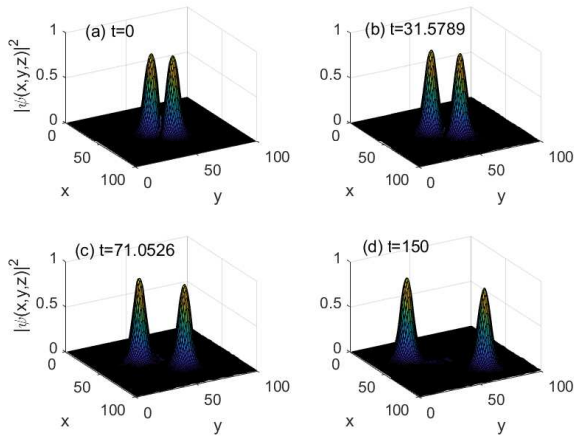}
\caption{(Color online) The repulsive interaction of two out-of-phase 3D
solitons, with phase difference $\protect\varphi =\protect\pi $, initial
distance $d=10$, and propagation constant $K=0.062$ of each soliton in the
initial state.}
\label{fig5}
\end{figure}

On the other hand, the interaction between out-of-phase fundamental 3D
solitons is repulsive \cite{potential}. In accordance with this expectation,
the simulations demonstrate, in Fig. \ref{fig5}, that the solitons with
phase shift $\varphi =\pi $ bounce back from each other, without any
conspicuous excitation of internal oscillations in either soliton.
Simulations of the 1D version of the NLSE with the CQ nonlinear terms also
reveal fast separation of the solitons (not shown here, as the result is
quite obvious).

\subsection{Symmetry breaking in the interaction of identical 3D solitons
with intermediate values of the phase difference.}

\begin{figure}[tbp]
\centering%{Sol3D_2x2L90T150LstPhasePi2Sft5Symm.eps}%
\includegraphics[width=0.99\columnwidth]{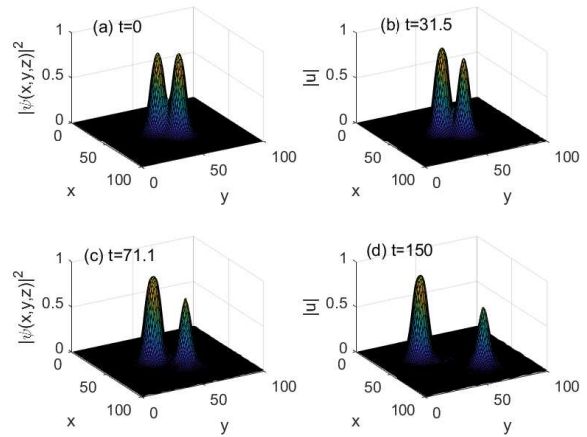}
\caption{(Color on line) The same as in Figs. \protect\ref{fig2} and \protect
\ref{fig5}, but for the 3D soliton pair with initial phase difference $%
\protect\varphi =\protect\pi /2$. The effective interaction is repulsive,
with conspicuous symmetry breaking of the solitons which bounce back from
each other.}
\label{fig6}
\end{figure}

\begin{figure}[tbp]
\centering%{Sol3D_isoL90T150LstPhasePi2Sft5Symm.eps}%
\includegraphics[width=0.99\columnwidth]{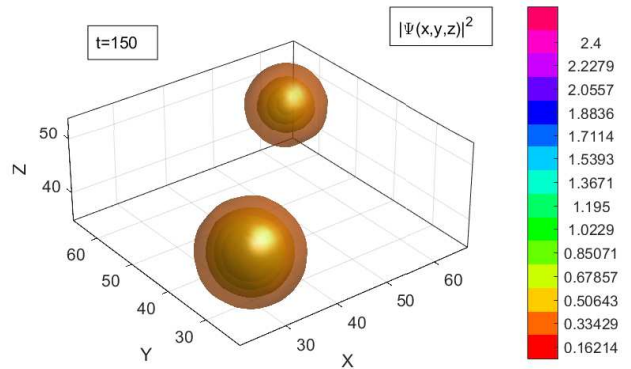}
\caption{(Color online) Shapes of the asymmetric solitons, observed in Fig.
\protect\ref{fig6} at $t=150$, are shown by isosurface of the local power.}
\label{fig7}
\end{figure}

The lowest-order approximation for the effective potential of the
interaction of fundamental 3D solitons gives zero for phase difference $%
\varphi =\pi /2$ between them \cite{potential}. On the other hand, direct
simulations, displayed in Fig. \ref{fig6}, reveal repulsion between the
solitons, which is coupled to the \textit{symmetry breaking}, that takes
place at the initial stage of the interaction: the separating solitons carry
obviously different energies (\ref{energy}), although input (\ref{input})
was composed of identical solitons. Further, Fig. \ref{fig7} corroborates
that the asymmetric solitons appear as almost unperturbed fundamental ones.

The explanation of the symmetry breaking in soliton-soliton collisions was
proposed (in terms of 1D solitons) in Ref. \cite{Lev}. Namely, the
interaction gives rise to opposite velocity vectors of the two solitons, $%
\pm \mathbf{v}$. It is well known that moving solitons in the NLSE acquire
the phase structure, represented by factors $\exp \left( \pm i\mathbf{v\cdot
r}\right) $ in the solutions, where $\mathbf{r}\equiv \left\{
x-x_{c},y-y_{c}\right\} $. Then, before the solitons, which were originally
introduced as per Eq. (\ref{input}), separate, their coherent superposition
gives rise to factor $\cos \left( \mathbf{v\cdot r}+\varphi \right) $. As
follows from here, there is a mismatch between the \textquotedblleft
amplitude center", located at $\mathbf{r}=0,z=0$, and the \textquotedblleft
phase center", which is shifted to $\mathbf{r}_{0}=-\varphi \mathbf{v}/v^{2}$%
. The mismatch qualitatively explains the breaking of the spatial symmetry
by the interaction between the solitons.

\begin{figure}[tbp]
\centering%{Sol3D_SSBL100t36PhiPi2V0Sft5PotOL_0All.eps}%
\includegraphics[width=0.99\columnwidth]{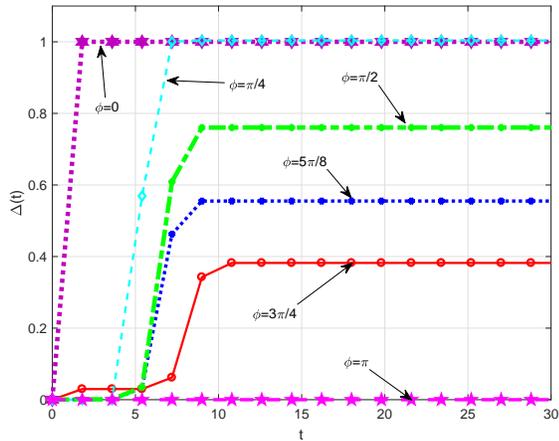}
\caption{(Color online) The symmetry-breaking measure, defined as per Eq. (%
\protect\ref{Delta}), is plotted versus $t$ for soliton-soliton interactions
at several values of the initial phase shift, $\protect\varphi $ (which
appears as $\protect\phi $ in this figure), between the initially identical
interacting solitons. The line with symbols, corresponding to $\protect%
\varphi =0$, represents the quick merger of in-phase solitons into a single
one, as shown in Fig. \protect\ref{fig1}, The line corresponding to  $%
\protect\varphi =\protect\pi /4$ also implies the merger, which occurs at $%
t\approx 7$. The merger formally corresponds to $\Delta =1$ (because the
single soliton is produced by the interaction), although the spatial
symmetry actually remains unbroken in this case. The interacting solitons
remain separated at $\protect\varphi >\protect\pi /4$. The out-of-phase
soliton pair ($\protect\varphi =\protect\pi $) separate without symmetry
breaking, keeping $\Delta \equiv 0$.}
\label{fig8}
\end{figure}

Results of the systematic study of this effect are summarized in Fig. \ref%
{fig8}, which displays evolution of the symmetry-breaking measure,
\begin{equation}
\Delta (t)\equiv \left\vert E_{1}(t)-E_{2}(t)\right\vert /\left[
E_{1}(t)+E_{2}(t)\right] ,  \label{Delta}
\end{equation}%
in the course of the collision, where $E_{1,2}(t)$ are energies (\ref{energy}%
) computed separately for the two solitons. This definition is formal for
the solitons merging into a single one (see Fig. \ref{fig2}), but it makes
sense in the case of the symmetry-breaking interaction because, as seen,
e.g., in Fig. \ref{fig6}, in such a case the solitons always remain
separated. In the limit of $t\rightarrow \infty $, $\Delta (t)$ takes a
final value corresponding to the pair of far separated solitons. As
explained in the caption to Fig. \ref{fig8}, $\Delta =1$ actually
corresponds to the merger of the pair into a single soliton, which does not
break the spatial symmetry [as mentioned above, definition (\ref{Delta})
does not adequately measure the symmetry breaking for merging solitons].
Therefore, the actual symmetry-breaking maximum in data displayed in Fig. %
\ref{fig8} may be identified as $\Delta \approx 0.76$, achieved at $\varphi
=\pi /2$. It is assumed that there is a critical value, $\varphi _{\mathrm{cr%
}}$, between $\varphi =\pi /4$ and $\pi /2$, at which the merger becomes
incomplete and a second soliton will appear in the outcome of the
interaction. Accurate identification of $\varphi _{\mathrm{cr}}$ is a
challenging objective for the 3D simulations.

\section{Rotating two-soliton molecules trapped in the external potential}

In the free space, the NLSE with the CQ nonlinearity does not produce
persistent bound states of 3D solitons (such as \textquotedblleft molecules"
or ring-shaped \textquotedblleft necklaces" \cite{cluster}). Here, we aim to
demonstrate that a rotating two-soliton \textquotedblleft molecule" may be
formed, in the 3D space, with the help of the 2D trapping potential (\ref{V}%
). To this end, centers of two identical solitons, with phase shift $\varphi
$ between them, were initially placed at points $\left( x-x_{c}=\pm
d/2,y-y_{c}=z=0\right) $, imparting to them initial velocities $\pm v_{0}$
in the direction of $y$ [i.e., multiplying the wave functions of individual
solitons by $\exp \left( \pm iv_{0}y\right) $].

\begin{figure}[tbp]
\centering%{Sol3D_3x3L100t450PhiPi2V03Sft5PotOL_AmpOL-1e-4-I.eps}%
\includegraphics[width=0.99\columnwidth]{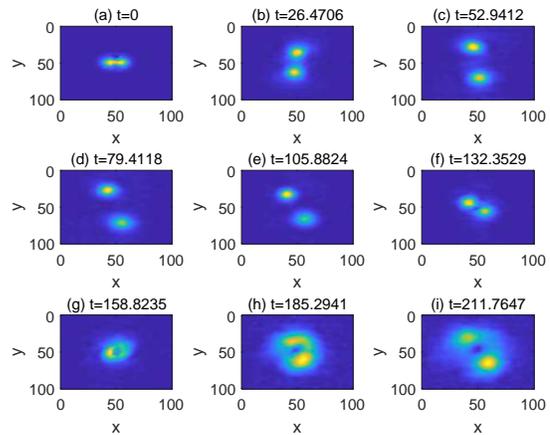}
\caption{(Color online) This figure and the following one display the
evolution of the soliton pair, set in the counter-clockwise quasi-stationary
rotational motion by initial velocities $\pm 0.3$ applied in the $y$
direction, perpendicular to the line connecting centers of the solitons,
separated by initial distance $d=10$, with phase shift $\protect\varphi =%
\protect\pi /2$ between them,\ in trapping potential (\protect\ref{V}) with
strength $V_{0}=10^{-4}$. The rotation period is $T\approx 150$.}
\label{fig9}
\end{figure}

\begin{figure}[tbp]
\centering%{Sol3D_3x3L100t450PhiPi2V03Sft5PotOL_AmpOL-1e-4-II.eps}%
\includegraphics[width=0.99\columnwidth]{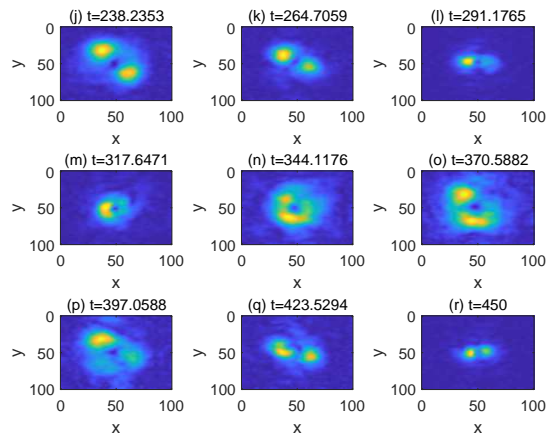}
\caption{The continuation of Fig. \protect\ref{fig9}.}
\label{fig10}
\end{figure}

In the absence of the trapping potential, the so built soliton pair in the
free space could perform the rotation by $180^{0}$ and would then break up
into separating solitons. On the other hand, Figs. \ref{fig9} and \ref{fig10}
demonstrate that, even a shallow trapping potential (\ref{V}), with strength
$V_{0}=10^{-4}$, helps to build a rotating ``molecule" with a
quasi-stationary form. This example is displayed for the soliton pair with
initial phase difference $\varphi =$ $\pi /2$ and separation $d=10$ along
the $x$ axis, the angular momentum being imparted by the transverse kick
with velocities $\pm 0.3$ in the $y$ direction. It is seen that, in
accordance to the sign of the kick, the soliton pair starts the rotation in
the counter-clockwise direction, with period $T\approx 150$. It is relevant
to mention that, as shown above, in physical units corresponding to typical
``light bullets", $T=150$ corresponds to the propagation distance $\sim 1$
cm in the bulk optical waveguide.

Figures \ref{fig9} and \ref{fig10} demonstrate three full rotations, in the
course of which the two solitons temporarily break their symmetry, in
agreement with the results reported in the previous section (see Fig. \ref%
{fig6}), but the symmetry is partly restored, from time to time. Similar
results were observed at other values of the parameters, although the
collection of systematic data is hampered by fact that these fully 3D
solitons are quite heavy. An alternative approach to the search of rotating
two-soliton bound states may be based on looking for stationary solutions to
Eq. (\ref{GPE-U}) with potential (\ref{V}) in a rotating reference frame.
Solution of this numerical problem is beyond the framework of the present
paper.

\section{Conclusion}

In this work, we aimed to perform the systematic numerical analysis of
interactions between identical 3D fundamental solitons in the NLSE
(nonlinear Schr\"{o}dinger equation) with the CQ nonlinearity, which is a
combination of self-focusing cubic and defocusing quintic terms. The model
applies to the spatiotemporal propagation in nonlinear optics and, under
certain conditions, to BEC. First, the interactions between two solitons
with phase shift $\varphi $ in the free space, separated by distance $d$,
were systematically simulated. The outcome of the interaction strongly
depends on $\varphi $, leading to the merger of two solitons into one at $%
\varphi =0$ (and at relatively small nonzero values of $\varphi $, such as $%
\varphi =\pi /4$), and the mutual rebound of the out-of-phase solitons at $%
\varphi =\pi $. A noteworthy finding is that the the in-phase fundamental
solitons merge into one in a nearly stationary form, without conspicuous
inner vibrations, unlike the results for the 1D version of the model, where
the solitons feature fusion into a strongly excited breather. At
intermediate values of the phase difference, such as $\varphi =\pi /2$, the
solitons separate, interacting repulsively, but, unlike the case of $\varphi
=\pi $, the interaction breaks the symmetry between the originally identical
solitons, producing two far separated ones with essentially different
energies (norms). The symmetry-breaking effect may be qualitatively
explained by means of an argument similar to one which was previously used
for the explanation of the symmetry breaking in the interaction of 1D
solitons \cite{Lev}. In the other part of the present work, it is
demonstrated that a pair of 3D solitons trapped in the 2D potential may form
a slowly rotating \textquotedblleft molecule", by initially kicking the
solitons in the opposite directions, perpendicular to the line connecting
their centers.

As an extension of the work, it may be interesting to systematically study
interactions of stable vortex solitons with topological charges $S_{1,2}=\pm
1$, which exist in the same 3D model \cite{9-authors}. In that case, the
result should depend, in particular, on the relative sign of $S_{1}$ and $%
S_{2}$. As mentioned above, it may also be relevant to develop a more
comprehensive analysis of the rotating bi-soliton \textquotedblleft
molecules" trapped in the external potential (in particular, looking for
them as stationary two-soliton complexes in the rotating reference frame).

\section*{Acknowledgments}

B.A.M. acknowledges partial support provided by the Binational (US-Israel)
Science Foundation through grant No. 2015616, and by Israel Science
Foundation (Grant No. 1287/17).

\end{document}